# Passive temperature management based on near-field heat transfer

**Sen Zhang, Wei Du, Wenjie Chen, Yongdi Dang, Naeem Iqbal and Yungui Ma***

State Key Lab of Modern Optical Instrumentation, Centre for Optical and Electromagnetic Research, College of Optical Science and Engineering; International research center for advanced photonics, Zhejiang University, China

*Corresponding author: yungui@zju.edu.cn

**Abstract**

Thermal or temperature management in modern machines has drawn great attentions in the last decades. The waste heat caused during the machine operation is particularly pernicious for the temperature-dependent electronics and may reduce the apparatus's performance and lifetime. To control the operation temperature while maintaining high input powers is often a dilemma. Enormous works have been done for the purpose. Here, a passive temperature management method based on near-field heat transfer is introduced, utilizing graphene-plasmon enhanced evanescence wave tunneling. Within a pump power tolerance range of 0.5-7 KW m$^{-2}$, the device can automatically regulate its thermal emissivity to quickly acquire thermal homeostasis around a designed temperature. It is compact, fully passive and could be incorporated into chip design.

The results pave a promising way for passive thermal management that could be used in modern instruments in particular for vacuum environmental applications.

# 1.Introduction

With the ever-increasing density of electronic components in a chip, thermal management has been an essential issue in device design as excessive heat accumulation may degrade the device performance or even threat their lifetimes. Classical heat exhausting methods mainly rely on conduction or convection, including miniature heat pipes (enhancing conductance),[1-3] piezoelectric fans (enhancing convection)[4-5] and transient phase change energy storage systems,[6] which have been vastly developed to control the temperature of powered electronic components. However, these approaches will consume additional energies and also often require large and separate accommodation space, undesired for compact, integrated and green device design.

With moderate heat input, alternatively, thermal radiation could be a good option to manage object's temperatures with a conformal coating layer usually consisting of paste of high thermal emissivity,[7-15] particularly advantageous for aerospace applications. The primary issue for far-field radiative cooling is the low output power compared with the traditional conduction or convection channels as it cannot exceed the blackbody radiation limit.[16-17] This obstacle could be partially overcome in the near-field scenarios,[18-23] where various potential usages have been explored.[24-27] When two objects are separated by a distance far smaller than the thermal wavelength,[28] thermally excited evanescent waves could transmit through the subwavelength gap and substantially enhance the heat flux. This effect could be strongly modulated between interfaces with large photonic density of state as such as between surfaces with surface plasmon polaritons (SPPs),[29] surface phonon polaritons (SPhPs)[30] or magnetic polaritons (MPs).[31]

In addition, semiconductor-based electronic or optoelectronic devices are usually designed to work within a desired temperature range. For them, a thermal homeostasis situation is important, often requiring a feedback to modulate the device's temperature. Compared with active methods mostly adopted, a fully passive self-adaptive temperature modulation has many vital advantageous on power consumption and space usage, but hard in practice as it requires materials with large temperature nonlinear emission properties. However, the recent vast development of radiative thermal diode incorporating phase change materials (PCMs) into photonic infrared metasurfaces gives a helpful clue to solve this issue.[32-35] Temperature-sensitive absorbers with high absorption contrast could be specifically designed when optical nano-antennas are modulated by a nearby PCM like vanadium dioxide ($VO_2$).

In this work, we propose a high-performance thermal homeostasis device that could passively control the temperature of heat pumped objects based on the scenario of near-field thermal radiation. The core part of the device consists of a thin (~ 2 μm) temperature-sensitive radiation coat made of $VO_2$ nanowire array (emitter and source side) and a graphene-boron nitride (BN) composite film (receiver and sink side), as shown in **Figure 1 a)**. The heat emitting and receiving components are separated by a nanometric vacuum distance. Under the assistance of the hybridized phonon-plasmon coupling, the near-field architecture will give rise to a radiation power significantly larger than the blackbody's limit with the orders dependent on the gap distance. This effect could be vastly modulated by the phase state of $VO_2$ (which has a metal-insulator phase transition around 341 K). With the nonlinear change of the $VO_2$'s dielectric property, the pumped object could quickly reach a thermal homeostasis around 333±7 K under the heat pump power varying in a wide range from 0.5 to 7 kW m$^{-2}$. The allowed input power range is more than one order larger than that for a far-field design.[12] The transient response could be modulated by optimizing the artificial structures and can be completed in sub-second time scales or even lower. In the end, important potential applications are envisioned for our compact and efficient heat management system.

**2. Device design**

Here, we managed to design a near-field thermal hysteresis device using a sort of metal insulator transition (MIT) material, $VO_2$, achieving a 0.5-7 kW m$^{-2}$ radiation power in nearly 14 K temperature scale. Vanadium dioxide ($VO_2$) is a phase change material which experiences insulator-metal phase transition at around 341 K.[36] As the phase change can also influence the optical properties,[37] we introduce this effect into near-field thermal management. **Figure 1 b)** describes the emissivity of insulating $VO_2$ and metallic $VO_2$ in far-field radiation, which clearly shows the phase transition in $VO_2$ can influence the emissivity as well. Considering the phase change of $VO_2$ will influence the optical properties and thus will influence the emissivity, we propose a structure that has a big variation in radiation capacity for metallic state or insulating state. To illustrate the principle, it is initially proposed for the temperature management of silicon based semiconductor chips. Figure 1 a) shows a scheme of our heat management system. From the top to bottom, it consists of the top Au layer for reflectance, BN layer covered by graphene, vacuum gap, VO$_2$ nanowire array (NWA), Au bottom layer and silicon substrate denoting the object to be thermally managed. We note that the emitter and the receiver are semi-infinite and the filling ratio of the nanowires is displayed veritably. The thickness of each layer denotes as d, $d_1$, $d_2$, $d_3$ for vacuum gap, VO$_2$ NW, BN and silicon, respectively. All the thicknesses are disproportionate and we zoom in some layers for better understanding. The thickness of Au layers is set at 200 nm that is thicker than the skin depth considered as non-transparent in infrared regime. We apply Au reflector on the back of BN and NWA to block the surface waves and gain a higher heat transfer power. BN layer and VO$_2$ nanowires are separated by a vacuum gap d=100 nm, and both have a thickness at 1μm. There is one sheet of graphene on the BN surface which can excite surface plasmon polaritons (SPPs). The first layer of the counterpart is needle-like VO$_2$ nanowires and the period of nanowires should be much smaller than the wavelength according to the effective medium theory.[38] The working part is designed of intrinsic silicon because of its wide usage in microelectronics. The thickness of silicon greatly influences the response time, therefore in this article, we switch the thickness in order to show wider practical applications. Cold side is set at 300 K constantly holding the same temperature as the ambient environment while the hot side changes around the phase transition temperature of VO$_2$. The heat flux generated by the working part

(silicon) conducts to VO2, raising the temperature then transfer to BN via radiation and thus dissipate into external environment via Au layer, which is depicted by red arrows in Figure 1 a). In conclusion, our passive temperature management device using metasurface and graphene-based structures to gain an enhancement in heat radiation.

As our proposed layer involved VO2 nanowire array structure, we utilize effective medium approximation (EMA) to obtain the effective dielectric functions and the expressions can be written as:[39]

$$\varepsilon_\parallel = 1 - f + \varepsilon_{\parallel,\mathrm{VO_2}} f \tag{1}$$

$$\varepsilon_\perp = \frac{\varepsilon_{\perp,\mathrm{VO_2}} + 1 + (\varepsilon_{\perp,\mathrm{VO_2}} - 1)f}{\varepsilon_{\perp,\mathrm{VO_2}} + 1 - (\varepsilon_{\perp,\mathrm{VO_2}} - 1)f} \tag{2}$$

$f$ is the filling ratio of nanowires, which can be 0~1 assuming the wire to be square. Where $\varepsilon_{\parallel,\mathrm{VO_2}}$ and $\varepsilon_{\perp,\mathrm{VO_2}}$ are the extraordinary and ordinary dielectric functions of VO2 (insulating state) uniaxial crystal, using the classical oscillator formula $\varepsilon(\omega) = \varepsilon_\infty + \sum_{i=1}^{N} \frac{S_i \omega_i^2}{\omega_i^2 - j\gamma_i\omega - \omega^2}$, $\varepsilon_\infty$ is the high-frequency constant, $S_i$ is the oscillator strength, $\omega_i$ is the phonon frequency, $j$ is the imaginary unit and $\gamma_i$ is the scattering rate. As for metallic state, Drude model can describe the dielectric function that is given by $\varepsilon(\omega) = \frac{-\omega_p^2 \varepsilon_\infty}{\omega^2 - j\omega\Gamma}$, all the parameters can be found in the article.[40] Boron nitride (BN) is isotropic and its dielectric function is given by $\varepsilon(\omega) = \varepsilon_\infty \frac{\omega^2 - \omega_{\mathrm{LO}}^2 + j\omega\gamma}{\omega^2 - \omega_{\mathrm{TO}}^2 + j\omega\gamma}$.[41] The dielectric functions of VO2 and BN are shown in **Figure 1 c)**. The effective dielectric functions for different filling ratios of VO2 nanowires are displayed in **Figure1 d)**. For the case when $f$=0.5 or 0.024 we find that the shapes of dielectric curves are similar but the values are relevant to the filling ratio, which indicates that we can switch the filling ratio to optimize the effective dielectric functions.

The mechanism of near-field radiation transfer in our structure can be comprehended as follows. Since the heat radiation power will be enhanced greatly by the SPPs[29] and SPhPs,[30] we design an emitter with $VO_2$ nanowires and a receiver made by graphene-covered BN. The metallic and insulating $VO_2$ can excite different surface polariton modes which provides strong couplings with graphene at metallic state and slightly enhances the radiation at insulating state. Graphene layer plays an important role in our structure as its excellent capability in near-field heat transfer. However, firstly in our design the emitter is covered by one sheet of graphene too, and at metallic state we do get greater enhancement but at insulating state the radiation power is still quite high, which lengthens the time to get a thermal homeostasis and influences the temperature range as well. Thus, we finally choose this asymmetry structure to avoid strong coupling when it is at insulating state. It is clear that our design is to enlarge the difference of radiation power between two states and enhance the radiation power at metallic state as large as possible to gain a better performance in input power tolerance and response time. Besides, the nanowires structure is studied[42] as its outstanding contributions in near-field heat transfer. In this article, we regulate the dielectric functions by changing the filling ratio of nanowires and use $VO_2$ nanowires to gain a larger radiation power. The optimization results discussed below.

**3.Calculations**

According to fluctuating electrodynamics and fluctuation dissipation theory, the net heat flux between two semi-infinite planes with their gap distance comparable to the radiation wavelength can be written as:[18,19,43]

$$P_{\text{rad}} = \frac{H'}{4\pi^2} \int_0^\infty d\omega [\Theta(\omega, T_2) - \Theta(\omega, T_1)] \tag{3}$$

$$H' = \frac{H_m}{2}\left[1 + \text{erf}\left(T_2 - T_c \pm \frac{w}{2}\right)\right] + \frac{H_i}{2}\left[1 - \text{erf}\left(T_2 - T_c \pm \frac{w}{2}\right)\right] \tag{4}$$

$T_1$ and $T_2$ are the temperature of the cold side and hot side, respectively. Where $\Theta(\omega, T)$ is the mean energy of thermal harmonic oscillators determined by $\Theta(\omega, T) = \hbar\omega/[\exp(\hbar\omega/k_\text{B}T) - 1]$, $\omega$ is the angular frequency. $H'$ is the rescaled heat transfer spectrum considering thermal hysteresis of VO$_2$. $erf$ is a build-in function in Matlab. $T_c$ is the midpoint of hysteresis range and $w$ is the width of thermal hysteresis range, in this article we assume them to be 333 K and 10 K if not particularly note. $H_{\text{m,i}}$ is the heat transfer spectrum for two phases (metallic and insulating) of vanadium dioxide given by:[44]

$$H_{\text{m/i}} = \int_0^\infty d\beta\beta [\tau_{\text{m/i,s}}(\omega, \beta) + \tau_{\text{m/i,p}}(\omega, \beta)] \tag{5}$$

and $\beta$ is parallel wavevector component. Furthermore, $\tau_\text{s}$ and $\tau_\text{p}$ are the energy transmission coefficients for s-polarized wave and p-polarized wave in the structure, which can be characterized as:[44]

$$\tau_\alpha = \begin{cases} \dfrac{(1 - |r_\alpha^c|^2)(1 - |r_\alpha^h|^2)}{|1 - \exp(2ik_0^z d) r_\alpha^c r_\alpha^h|^2}, & \beta < \dfrac{\omega}{c} \\ \dfrac{4\text{Im}(r_\alpha^c)\text{Im}(r_\alpha^h) \exp(-2|k_0^z|d)}{|1 - r_\alpha^c r_\alpha^h \exp(-2|k_0^z|d)|^2}, & \beta > \dfrac{\omega}{c} \end{cases} \tag{6}$$

subscript $\alpha = \text{s}, \text{p}$ represents different polarization of light and Im denotes the imaginary part. $k_0^z = \sqrt{k_0^2 - \beta^2}$ is the vertical component of wave factor in vacuum

($k_0$). $d$ is the gap distance between the two sides. $r_\alpha^c$ and $r_\alpha^h$ are the reflection coefficients of α-polarized for incident light from vacuum to cold side or hot side. Note that in our case, both cold side and hot side have a multilayer structure, which can be expressed as:

$$r_p^{0i} = \frac{r_p^{01} + (1 - r_p^{01} - r_p^{10})r_p^{1i} \exp(2ik_1^z d_1)}{1 - r_p^{10} r_p^{1i} \exp(2ik_1^z d)} \tag{7}$$

$$r_p^{1i} = \frac{r_p^{12} + (1 - r_p^{12} - r_p^{21})r_p^{2i} \exp(2ik_2^z d_2)}{1 - r_p^{21} r_p^{2i} \exp(2ik_2^z d_2)} \tag{8}$$

$$\vdots$$

$$r_p^{(i-2)i} = \frac{r_p^{(i-2)(i-1)} + \left(1 - r_p^{(i-2)(i-1)} - r_p^{(i-1)(i-2)}\right) r_p^{(i-1)i} \exp(2ik_{i-1}^z d_{i-1})}{1 - r_p^{(i-1)(i-2)} r_p^{(i-1)i} \exp(2ik_{i-1}^z d_{i-1})} \tag{9}$$

and for s-polarization:

$$r_s^{0i} = \frac{r_s^{01} + (1 + r_s^{01} + r_s^{10})r_s^{1i} \exp(2ik_1^z d_1)}{1 - r_s^{10} r_s^{1i} \exp(2ik_1^z d)} \tag{10}$$

$$r_s^{1i} = \frac{r_s^{12} + (1 + r_s^{12} + r_s^{21})r_s^{2i} \exp(2ik_2^z d_2)}{1 - r_s^{21} r_s^{2i} \exp(2ik_2^z d_2)} \tag{11}$$

$$\vdots$$

$$r_s^{(i-2)i} = \frac{r_s^{(i-2)(i-1)} + \left(1 + r_s^{(i-2)(i-1)} + r_s^{(i-1)(i-2)}\right) r_s^{(i-1)i} \exp(2ik_{i-1}^z d_{i-1})}{1 - r_s^{(i-1)(i-2)} r_s^{(i-1)i} \exp(2ik_{i-1}^z d_{i-1})} \tag{12}$$

$r_p^{0i}$ or $r_s^{0i}$ ($i = 2,3,4 \ldots$, is the total of layers in the structure) is the reflection coefficient from layer $i - 2$ to layer i. $d_{1,2,3\ldots i-1}$ and $k_{1,2,3\ldots i-1}^z$ are the thickness and vertical component of wave factor for each ($1, 2, 3 \ldots i - 1$) layer, respectively. According to Fresnel's equations, the reflection coefficient from A to B can be written as:[45]

$$r_p^{AB} = \frac{k_A^z \varepsilon_B - k_B^z \varepsilon_A}{k_A^z \varepsilon_B + k_B^z \varepsilon_A} \tag{13}$$

$$r_s^{AB} = \frac{k_A^z - k_B^z}{k_A^z + k_B^z} \tag{14}$$

if the interface is covered by graphene:[45]

$$r_p^{AB} = \frac{k_A^z \varepsilon_B - k_B^z \varepsilon_A + \frac{\sigma k_A^z k_B^z}{\omega \varepsilon_0}}{k_A^z \varepsilon_B + k_B^z \varepsilon_A + \frac{\sigma k_A^z k_B^z}{\omega \varepsilon_0}} \tag{15}$$

$$r_s^{AB} = \frac{k_A^z - k_B^z + \sigma \mu_0 \omega}{k_A^z + k_B^z - \sigma \mu_0 \omega} \tag{16}$$

where $\varepsilon_0$ and $\mu_0$ are the vacuum permittivity and permeability, $\varepsilon_{A/B}$ is the dielectric function of medium A or B. The vertical wave factor can be written as $k_{A/B}^z = \sqrt{\varepsilon_{A/B} k_0^2 - \beta^2}$ but if the medium is uniaxial crystal, $k_{A/B}^z = \sqrt{\varepsilon_{\perp,A/B} k_0^2 - \varepsilon_{\perp,A/B}/\varepsilon_{\parallel,A/B} \beta^2}$ for p-polarization and $k_{A/B}^z = \sqrt{\varepsilon_{\perp,A/B} k_0^2 - \beta^2}$ for s-polarization, where the subscript "⊥" or "∥" represents different dielectric function in different directions(vertical or parallel to the optic axis). $\sigma$ is the conductivity of graphene, which can be determined as $\sigma = \frac{ie^2 |E_F|}{\pi \hbar^2 (\omega + i\tau_g^{-1})}$.[46]

To demonstrate the time-domain response, the time is related to the whole thermal capacity of our device:[12]

$$\frac{dT}{dt} = \frac{P_{rad} - P_{in}}{\rho l c} \tag{17}$$

$\rho$ is the density (kg m$^{-3}$) $l$ is thickness (m) and $c$ is the specific heat capacity (J K$^{-1}$ kg$^{-1}$) of the hot side. **Equation 17** indicates that the time response is determined by the properties of materials and the radiation power of our thermal hysteresis device. Since the silicon layer is much thicker than any other materials, the influence of VO$_2$ and Au become negligible in our case, thus we assume $\rho = 2328.3$ kg m$^{-3}$, $l = 10^{-4}$ m and $c = 703$ J K$^{-1}$ kg$^{-1}$ which are the properties of intrinsic silicon.[47]

We discuss the optimization of two parameters of VO$_2$ nanowires structure in this section. The gap distance or other conditions that can influence the radiation flux are constant while we doing the calculations. Filling ratio of nanowires, as calculated in **Equation 1** and **2**, is the ratio of nanowires area occupied the whole area, considering the nanowire to be cuboid. In particular, when filling ratio is 1 that means nanowires become thin film structure in terms of the EMA theory. Next tunable parameter is the height of nanowires. We create a new function, $Q_a$, which can be written as:

$$Q_a = \frac{(P_m - P_i)^2}{P_m + P_i} \tag{18}$$

where $P_m$ and $P_i$ are the radiation power of metallic and insulating $VO_2$ at 358 K and 318 K. The temperature is chosen to make sure the $VO_2$ is in metallic or insulating state. The unit of $Q_a$ can be W m$^{-2}$. This objective function is an evaluation of the difference between the two states, thus the optimization results are contingent and may not get the maximum or minimal value of the net heat flux for our structure. As shown in **Figure 2**, we plot four solid lines and two dashed lines with different thicknesses and the exact values are given in the legend. The heights are increasing from bottom to up curves and the dashed lines are almost same. It is clear that the height of $VO_2$ nanowires determines the upper limit of our results. For practical concern, 1 μm height is the best choice for our design. Besides, as the heights increasing, the optimized filling ratio switches from 0.231 (10 nm height) to 0.024 (1 μm height) indicating the significance of effective dielectric functions matching in different situations. Moreover, the Fermi level of graphene may also change the optimization results but we are more concerned about the properties of hot side layers, so it will not be discussed herein.

## 4. Results

To figure out the difference of radiation power mentioned above; we next calculate the energy transmission coefficients of three structures based on the $VO_2$ nanowires without graphene covered the cold side, $VO_2$ thin film (same height to the nanowires) with graphene covered cold side and $VO_2$ nanowires with graphene covered cold side. We only calculate the p-polarized coefficients neglecting the s-polarized part due to the negligible contributions of s-polarized wave in near-field regime. As shown in the **Figure 3**, the metallic state supports surface plasmon polaritons while the insulating state supports surface phonon polaritons. Importantly, comparing **Figure 3 e)** and **f)**, the SPPs excited by graphene have a strong coupling with metallic $VO_2$ nanowires rather than insulating nanowires, explaining the big difference in radiation abilities. Synchronously, **Figure 1 c)** and **d)** illustrate that the phonon frequency of BN (around $2 \times 10^{14}$ rad s$^{-1}$) mismatches that of $VO_2$ (around $0.5 \times 10^{14}$ rad s$^{-1}$), so that the BN will not couple well with the insulating $VO_2$, making it possible to design a device that generate less radiation heat in insulating state. **Figure 3 a)** and e) indicate the importance of graphene and explain the weak heat transfer between the nanowires and bare BN. **Figure 3 c)** and **d)** show that $VO_2$ thin film does not have a strong coupling with graphene in both phase states under this condition and thus generate less heat radiation. What's more, we find that the SPPs excited by higher Fermi level can match with the $VO_2$ thin film though it's not shown in the article.

In **Figure 4** we calculate the net heat flux of four structures for comparison, except three structures discussed above, we add a curve plotting the heat flux between two black bodies. The X-axis represents the temperature of the hot side. All the structures hold the same gap distance and temperature range so that we can clearly notice the superiority of our designs. Firstly, we compare the black bodies with the other three conditions. It indicates that both the nanowires and thin film with the graphene structures exceed the blackbody's limit. In the near-field regime graphene will strongly enhance the coupling between the BN and the $VO_2$ with the coupling of the SPPs and SPhPs. Comparing the results of thin film (blue lines) and the nanowires (red lines), the differential of nanowires with graphene between metallic state and insulating state is much larger, denoting the necessity of nanowires in our device. Besides, the change in radiation power become tremendous during the phase transition gives us the opportunity to design a quick response device using the $VO_2$ nanowires. We choose a range at 523.7 W m$^{-2}$ ($T_2$ at 326.1 K) -6959 W m$^{-2}$ ($T_2$ at 339.8 K) for input power, in other words, it is the height of thermal hysteresis loop so that our device can reach the thermal homeostasis as quick as possible under the premise of a large put-in power, shown in **Equation 19**. Nearly 13.7 times of difference in put-in power can be applied to our device, giving an extensive usage in thermal management. In addition, the filling ratio used in our calculations is optimized at 0.024, which has been discussed in optimization section.

    The analysis of time-domain response is an essential part of modern thermal management. Generally, the heat dissipation by radiative means may not be as quick as the conductive ways, but the relaxation time is still acceptable according to our results. As for our device shown in Figure1 a), we assume a silicon layer of 100 μm

to be the working part on the bottom of the structures which is also the input heat source. The function of input energy power is given as follows:

$$P_{in}(t) = \begin{cases} P_{rad}(339.8 \text{ K}), & t \leq \dfrac{T_w}{2} \\ P_{rad}(326.1 \text{ K}), & t > \dfrac{T_w}{2} \end{cases} \quad (19)$$

where the $t$ is the actual time and $T_w$ is the whole time. Since other parts of our device (VO$_2$ nanowires, BN, Au and graphene) are very thin in vertical scale compared with silicon, we assume the parameter in **Equation 17** can be simply given as the parameters of silicon. During the calculations we find that changing input power will have great impact on response time. **Figure 5 a)** depicts only the proper value of input power range that response quickest. When we increase the crest value of input power, in general way, the response should be quicker, but as the temperature surpassing 339.8 K (when phase transition completed) the radiation power changes slow, requiring a higher $T_2$ to get to the thermal homeostasis, thus the working temperature will exceed our design limit. On the other hand, the $P_{rad}$ increases slowly after completing transition, but $P_{in}$ is still quite high, explaining the red line in Figure 5 a) cannot achieve to a steady state ($P_{in}=P_{rad}$) in two seconds. In comparison, we plot dashed lines (green and yellow) which is within the proper range, indicating our device is suitable for these situations. The temperature varying is from 325 K to 340 K for input power within the proper range, and for that exceeds the range, the temperature difference become large as well. We further demonstrate the width of temperature range for thermal hysteresis can influence the thermal homeostasis temperature at the same input power. We plot four lines in **Figure 5 b)** and numerically set the width as 4 K to 14 K while the calculations above were done at 10 K, and the input power is set at $P_{in} = 6959/523.7$ W m$^{-2}$. The results show that the width and temperature range during heating-up and cooling-down process obey positive correlation, which we can take advantage of to implement smaller temperature change for a temperature-sensitive machine. With width decreasing, the responses of our device become slow (the red and yellow dashed lines) owing to the width changes the transition temperature hence the net power ($P_{in} - P_{rad}$) reduces at the same temperature. Some experimental articles point out the hysteresis width can be switched[48-50] which pave the way for tunable devices in the future. **Figure 6 a)** changes the thickness of silicon layer, which demonstrates the ability of temperature management for thicker working part. Even when the thickness increases to 1 mm, our device can achieve the thermal homeostasis in 5 seconds. **Figure 6 b)** shows that the uncoated silicon cannot achieve thermal homeostasis in four seconds, and the working temperature keeps rising even during the through of the input power. The VO$_2$ nanowires not only decrease the response time but also shrink the span of temperature during a heating and cooling process, making it possible to implement such a device for thermal management. In summary, our device is designed by several thin layers and is capable of thermal homeostasis under nearly 7 KW m$^{-2}$ input power, in the meanwhile control the temperature change and response time into an acceptable range.

**5.conclusion**

In summary, we have demonstrated a fancy design of near-field thermal homeostasis device mainly capitalizing the phase change of VO$_2$ to gain a 13.7 times difference in radiation flux between the two states. We also calculate the response time of our device, and it is indicated that within a proper range of input power, our device could get to the homeostasis in 0.5 seconds no matter it is a heating or a cooling process. We switch the input power to test the power bearing capacity and we decrease the width of hysteresis loop finding a narrower temperature range during the process to a steady state under the same input power. Besides, we compare the uncoated silicon with coated one to examine the remarkable impact of our device in shrinking the temperature variation and decreasing the response time. We numerically calculate the optimization of VO$_2$ nanowires' parameters including thickness and filling ratio via EMA theory and an objective function $Q_a$ to evaluate where the maximum difference is. There are some other elements like the Fermi level of graphene, the structure of BN and other near-field efficient materials or metasurfaces that can be studied to further increase the performance.

((Insert Figure here. Note: Please do not combine figure and caption in a textbox or frame.))

**Figure 1.** a) Schematic of a near-field thermal homeostasis device. Upper side (cold side) consists of BN covered graphene and Au layers, which holds a temperature at $T_1$=300 K. Lower side (hot side) consists of VO$_2$ nanowires, Au and Si (as a working object), which holds a temperature at $T_2$. The whole device is in vacuum. The red

arrow is the direction of heat flux.

b) Emissivity of VO$_2$ in the far-field region. The results are calculated based on VO$_2$ bulk structures.

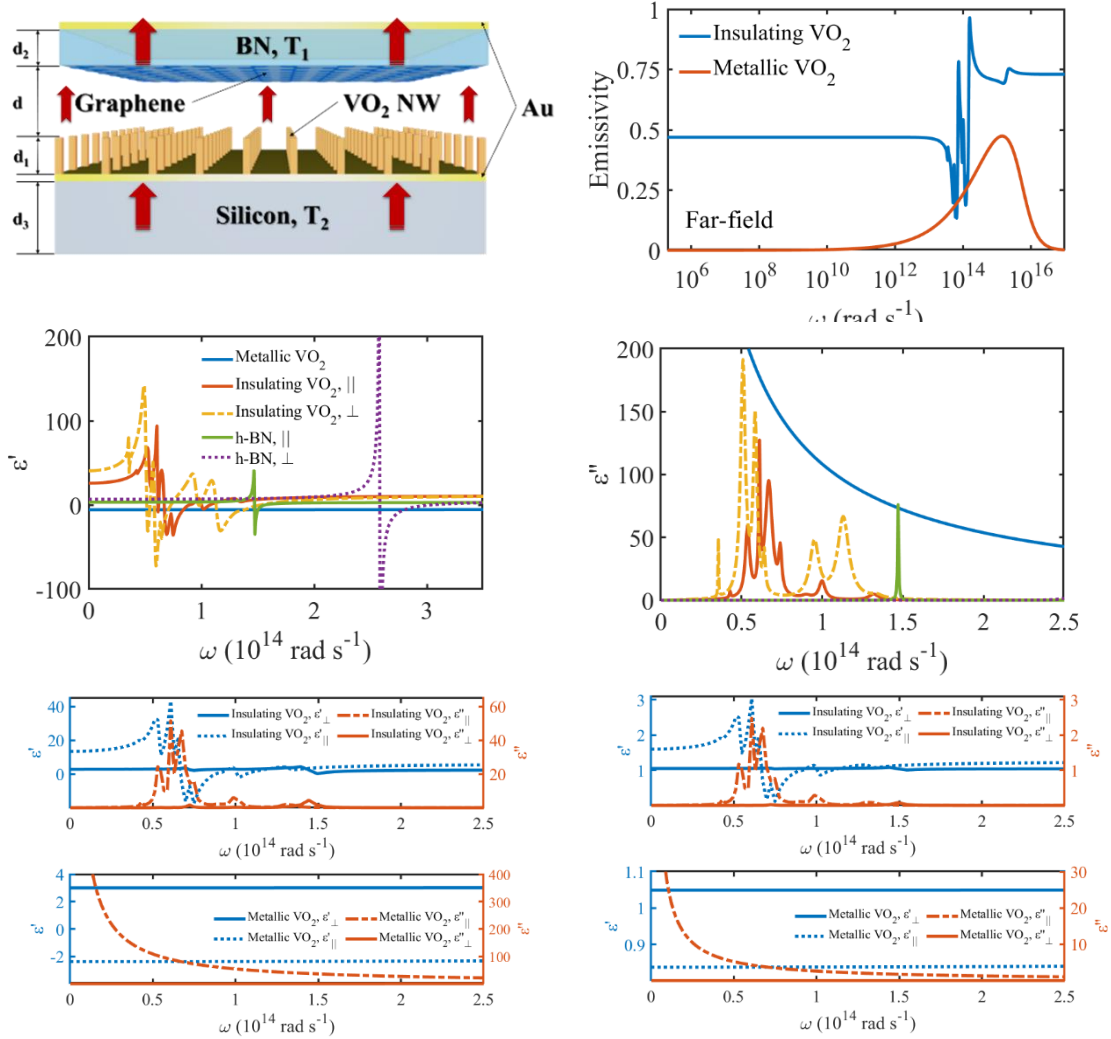

Dielectric functions for metallic and insulating VO$_2$ and BN. $\varepsilon'$ is real part c) and $\varepsilon''$ is imaginary part d). The perpendicular and parallel dielectric functions for insulating VO$_2$ are the yellow and red lines. The blue lines are for metallic VO$_2$ and the dashed purple lines are for BN.

Plot of effective dielectric functions for $f=0.5$ e) and $f=0.02$ f). In each inset, $\varepsilon'$ (left Y axis) is real part and $\varepsilon''$ (right Y axis) is imaginary part. Solid lines represent the vertical effective dielectric functions while dashed lines represent parallel functions.

**Figure 2.** Optimization of nanowires structures. X axis represents the filling ratio of VO₂ nanowires, Y axis, $Q_a$ is an objective function characterizes the difference of metallic VO₂ and insulating VO₂. Several discrete thicknesses of nanowires are plotted in different color, the solid lines are bottom-up as 10 nm, 40 nm, 70 nm and 100 nm, the light blue dashed line is 1 μm and the green dashed line is 10 μm.

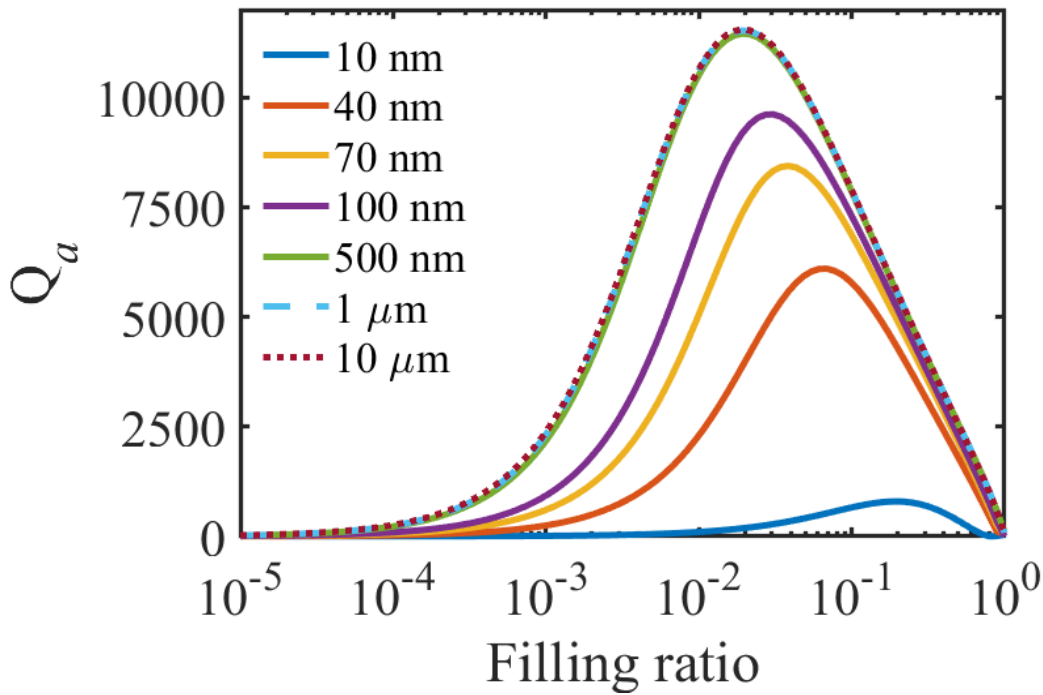

**Figure 3.** Energy transmission coefficients for p-polarization of three structures against in-plane wave factor $\beta$ and angular frequency $\omega$. a), c), e) are for metallic phase of VO₂ and b), d), f) are for the insulating ones. All calculations are done under a gap distance at 100 nm, cold side temperature at 300 K and hot side at 350 K. For graphene, the Fermi level is set at 0.1 eV. The thickness of $VO_2$ (nanowires or thin film) or h-BN film is 500 nm and the filling ratio is 0.02.

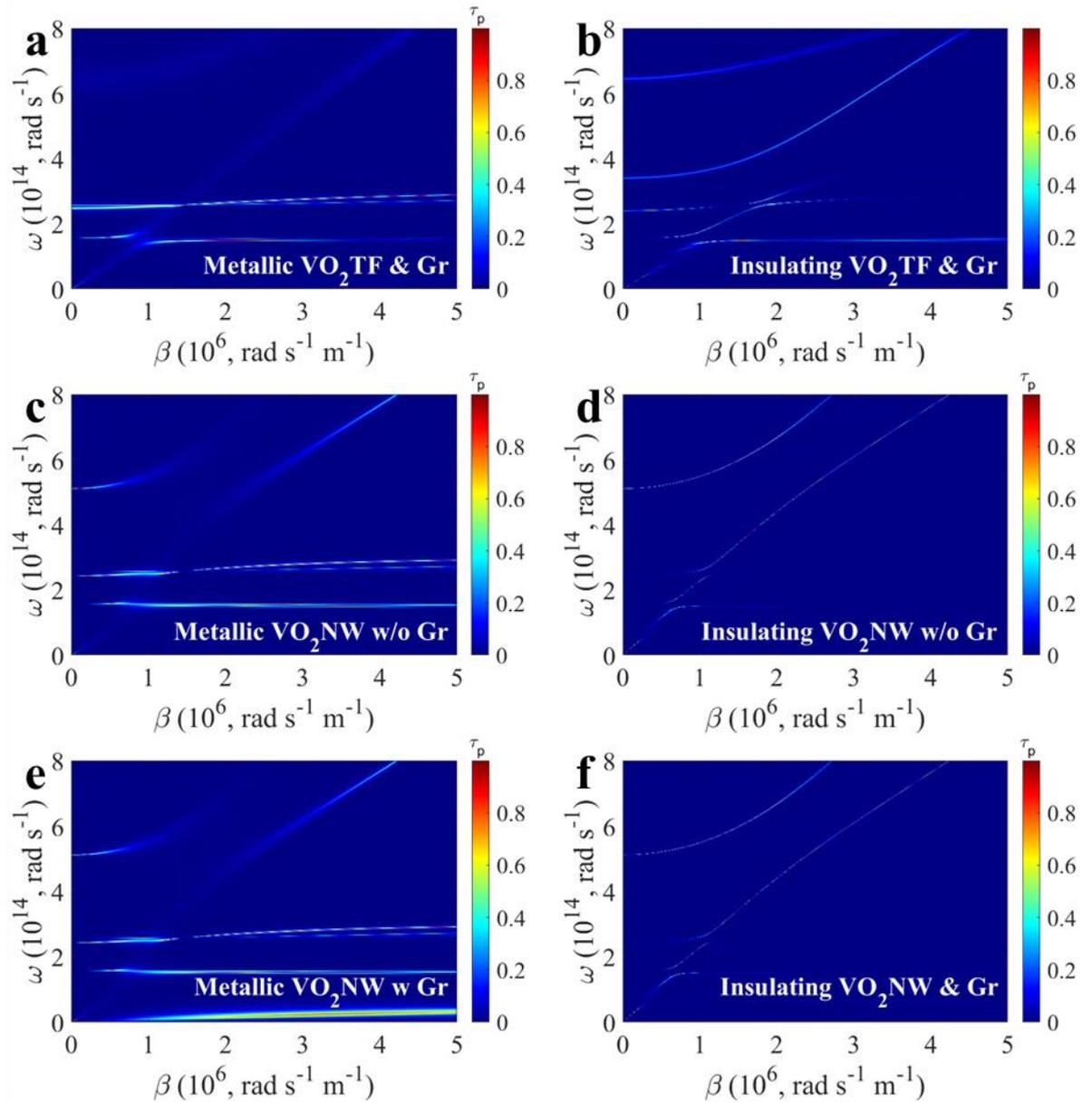

**Figure 4.** Thermal hysteresis loop. The solid lines denote heating-up procedure and colored dashed lines denote cooling-down procedure. This figure compares blackbody (black dashed line) with three different structures including VO₂ nanowires and graphene covered BN (VO2 NW & Gr), VO2 nanowires and without graphene covered BN (VO2 NW w/o Gr) and VO2 thin film and graphene covered BN (VO2 TF & Gr).

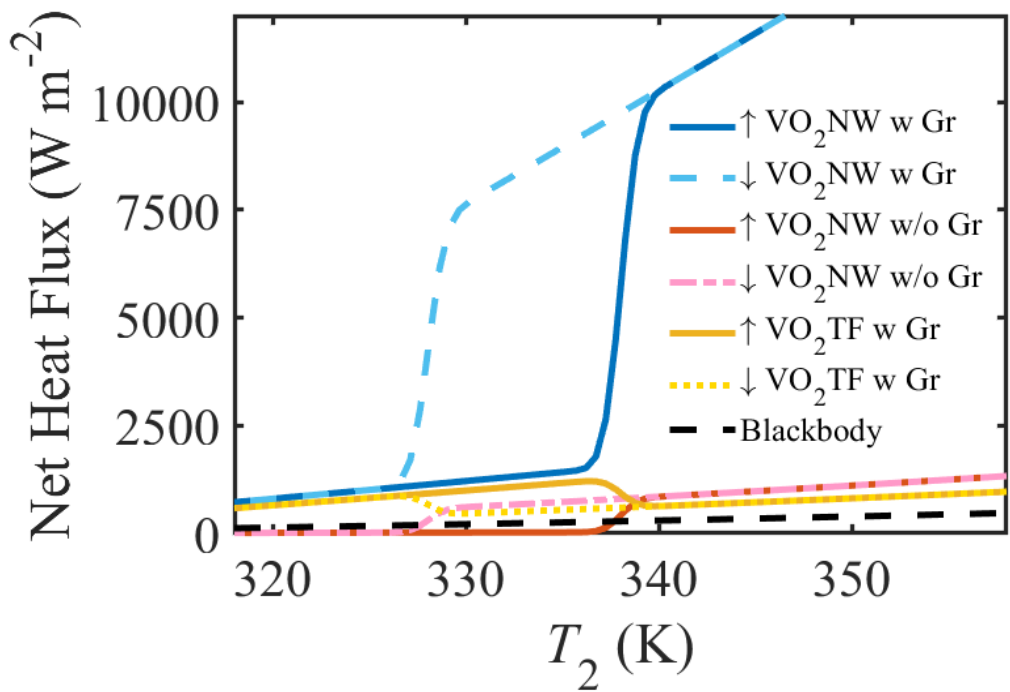

**Figure 5.** Response time for a) different $P_{in}$ and b) different hysteresis width of VO₂. $P_{in}$ is set as a square wave input power which is at crest in 0-2 seconds to heat the silicon up and at through in 2-4 seconds to cool it down. The width of thermal hysteresis range is set at 4, 6, 10, 14 K, respectively.

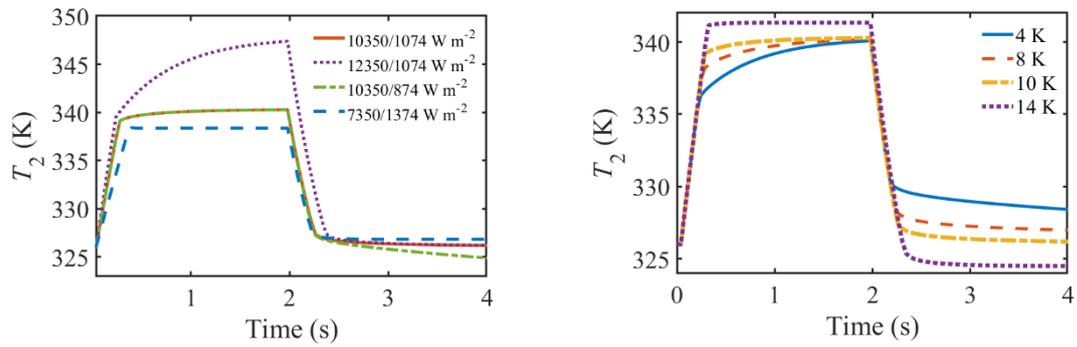

**Figure 6.** Time response for a) different thicknesses of silicon layer and b) bare silicon compared with our temperature management device. The figure is plotted under the same conditions, such as input power, gap distance, structure of sink side et

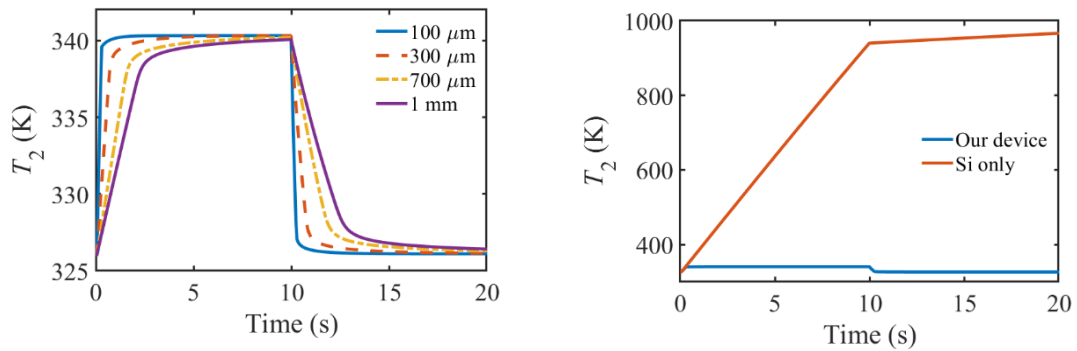

al.

**Supporting Information** ((delete if not applicable))
Supporting Information is available from the Wiley Online Library or from the author.

**Acknowledgements**
((Acknowledgements, general annotations, funding. Other references to the title/authors can also appear here, such as "Author 1 and Author 2 contributed equally to this work."))